\documentclass[12pt,useAMS,usenatbib]{mn2e}

\usepackage{color}
\usepackage{graphicx}
\usepackage{aas_macros}
\usepackage{times}
\usepackage{amsmath,amsfonts,amssymb}

\def\lesssim{\mathrel{\hbox{\rlap{\hbox{\lower4pt\hbox{$\sim$}}}\hbox{$<$}}}}
\def\gtrsim{\mathrel{\hbox{\rlap{\hbox{\lower4pt\hbox{$\sim$}}}\hbox{$>$}}}}

\title[Stochastic Star Formation \& Feedback in Low-Mass Galaxies]
      {Stochastic Star Formation \& Feedback: Mapping Low-Mass Galaxies to 
        Dark Matter Haloes}
      \author[C. Power, G. A. Wynn,  A. Robotham, G. F. Lewis, M. I. Wilkinson]
	     { C. Power$^{1}$\thanks{chris.power@icrar.org},
               G. A. Wynn$^{2}$, A. Robotham$^{1}$, G. F. Lewis$^3$ \& 
               M. I. Wilkinson$^{2}$\\
               $^1$ International Centre for Radio Astronomy Research,
               University of Western Australia, 35 Stirling Highway, Crawley, 
               WA 6009, Australia\\
               $^2$ Department of Physics \& Astronomy, University of 
               Leicester, Leicester LE1 7RH, United Kingdom\\
               $^3$ Sydney Institute for Astronomy, School of Physics, 
               A28, The University of Sydney, NSW 2006, Australia}
             
\begin{document}

\date{}

\pagerange{\pageref{firstpage}--\pageref{lastpage}} \pubyear{2014}

\maketitle

\label{firstpage}

\begin{abstract}
 Comparison of observed satellite galaxies of the Milky Way (hereafter MW)
 with dark matter subhaloes in cosmological $N$-body simulations of MW-mass 
 haloes suggest that such subhaloes, if they exist, are occupied by satellites in a
 stochastic fashion. We examine how inefficient massive star formation and associated
 supernova feedback in high-redshift progenitors of present-day low-mass subhaloes 
 might contribute to this stochasticity. Using a Monte Carlo approach to follow
 the assembly histories of present-day low-mass haloes with 
 $10^7 \lesssim M \leq 10^{10}$ ${\rm M}_{\odot}$, we identify when cooling and star 
 formation is likely to proceed, and observe that haloes with present-day masses 
 $\lesssim 10^9 {\rm M}_{\odot}$ never grow sufficiently massive to support atomic 
 hydrogen line cooling. Noting that the star formation timescale decreases 
 sharply with stellar mass as $t_{\rm PMS} \propto m_{\ast}^{-2.5}$, we argue that, 
 should the conditions for high mass star formation arise in low-mass haloes, the ensuing 
 supernovae are likely to disrupt ongoing lower-mass star formation and unbind gas within the 
 halo. This potentially star-forming gas is unlikely to be replenished in lower mass haloes 
 because of, e.g. cosmological reionization, and so we expect galaxy formation to be 
 stymied in a manner that depends on host halo assembly history and the efficiency and 
 timing of star formation in proto-galaxies, which we illustrate using a Monte Carlo model. 
 Based on these simple physical arguments, we assert that stochasticity of star 
 formation and feedback is an essential but overlooked ingredient in modelling galaxy 
 formation on the smallest scales.
\end{abstract}

\begin{keywords}
  galaxies: formation -- galaxies: evolution -- stars: formation -- 
  methods: analytical -- methods: numerical

\end{keywords}

\section{Introduction}
\label{sec:intro}

A defining prediction of the Cold Dark Matter (hereafter CDM) model
of structure formation is that dark matter haloes contain a wealth of substructure 
(hereafter subhaloes) whose abundance increases with decreasing mass $M$, approximately 
as $M^{-0.8}$ \citep[e.g.][]{springel.etal.2008}. The connection between this 
subhalo population and the hosts of galaxies in galaxy clusters has been
appreciated for some time \citep[e.g.][]{white.etal.1987,frenk.etal.1988},
but it required the emergence of cosmological $N$-body simulations with 
sufficiently high mass and force resolution to overcome overmerging 
\citep[cf.][]{moore.etal.1996,klypin.etal.1999a} to establish that the presence
of substructure in dark matter haloes is a presistent feature of the CDM model.
It was soon appreciated that this signals a problem on the scale of galaxies.
If subhaloes in cluster-mass systems are the hosts of cluster galaxies, then
subhaloes in galaxy-mass systems should host satellite galaxies and we should
expect many more satellites around galaxies like the Milky Way and Andromeda 
than are observed -- the so-called ``Missing Satellites'' problem 
\citep[e.g.][]{klypin.etal.1999b,moore.etal.1999}. 

This apparent tension between simulation and observation has led to over a decade of 
investigation in which both astrophysical (e.g. feedback from supernovae, cf. 
\citealt{dekel.silk.1986}, or from a photo-ionizing background during cosmological 
reionization, cf. \citealt{efstathiou.1992}) and particle physics (e.g. revision of the 
dark matter model) explanations for the apparent dearth of satellite galaxies around 
Milky Way-type galaxies have been explored. The challenge remains to construct
a physically-motivated predictive theory to map the subhalo population identified in 
cosmological $N$-body simulations to a plausible realisation of the observed satellite 
galaxy population of the Milky Way. Despite this, empirically matching the abundance of 
observed satellites with subhaloes identified in cosmological $N$-boy simulations has led 
to the conclusion that satellite galaxies occupy their dark matter (sub)haloes in a 
stochastic fashion; that is, a simple linear mapping of satellite galaxy to dark matter 
host based on, for example, subhalo mass, cannot reproduce the observed abundance and 
spatial distribution of satellite galaxies \citep[cf.][]{boylan.etal.2011}. The question remains as to what drives this 
stochasticity? 

The environmental dependence of cosmological reionization offers one possibility.
Low-mass haloes that formed sufficiently early, prior to reionization, could have formed 
stars, and it is likely that these correspond to the satellites that we see today 
\citep[e.g.][]{moore.etal.2006}. For example, the number density of dwarf galaxies in 
galaxy clusters is much higher than in the field, reflecting the biased nature of galaxy 
formation -- dwarf galaxies in overdense regions would have formed earlier and in greater 
numbers than their counterparts in mean- or low-density regions. Reionization would have 
had the effect of spatially filtering the abundance of satellite galaxies -- that is, more 
dwarfs would have formed in overdense regions than in mean- or low-density regions 
\citep[e.g.][]{weinmann.etal.2007}. However, the spatial scales over which this
filtering would have occurred suggests that this form of suppression cannot be considered 
stochastic.

The stochasticity of star formation and consequent feedback in low-mass haloes offers another 
possibility. Recent observational evidence \citep[e.g.][]{skillman.etal.2014,weisz.etal.2014} 
and theoretical considerations \citep[e.g.][]{madau.etal.2014,kelson.2014} suggests that star 
formation is likely to have proceeded in a stochastic fashion in low-mass systems. We are 
particularly interested in the consequences of stochasticity in the rate at which high mass 
stars form and the resulting feedback (e.g. photo-ionization, stellar winds, supernovae); in 
this short paper, we focus on the impact of supernovae in these systems. Simple physical 
arguments, which we set out in \S\ref{sec:theory}, suggest that star formation should be 
inefficient in low-mass haloes and that scaling relations such as 
Schmidt-Kennicutt are likely to break down at low masses. This implies that the likelihood 
of stars massive enough to result in supernovae ($M \gtrsim 8 {\rm M}_{\odot}$) is low. 
However, when they do form, they form rapidly -- the pre-Main Sequence timescale 
($t_{\rm PMS}$) declines sharply with increasing stellar mass as 
$t_{\rm PMS} \propto m_{\ast}^{-2.5}$ \citep[cf.][]{2013sse..book.....K}, and so 
high mass stars may reach the Main Sequence and evolve along it to 
yield supernovae on a timescale that is shorter than that required for lower-mass stars 
($\sim \rm M_{\odot}$) to form. The injection of $\sim 10^{51}$ erg of mechanical energy per
supernovae into the IGM can potentially unbind the gas in the galaxy, which occurs on mass 
scales of order $10^7-10^8 {\rm M}_{\odot}$ for single massive stars. We investigate this 
picture using Monte Carlo halo mass assembly histories and star formation models 
(cf. \S\ref{sec:model}) and discuss
how it can help to explain the low observed occupancy rate of low-mass dark matter 
subhaloes with satellites, which we summarise in \S\ref{sec:summary}.

\section{Stochastic Star Formation \& Feedback in Low-Mass Haloes: Analytical Framework}
\label{sec:theory}

We consider a dark matter halo, which we treat as a spherically overdense region
of virial mass $M_{\rm vir}$, 
\begin{equation}
  M_{\rm vir} = \frac{4\pi}{3}\,\Delta_{\rm vir}\,\rho_{\rm crit}\,r_{\rm vir}^3
\end{equation}
where $\Delta_{\rm vir}$ is the virial overdensity criterion,
$\rho_{\rm crit}=3H^2/8\pi\,G$ is the critical density of the Universe at redshift $z$, 
$H$ is the Hubble parameter
and $G$ the gravitational constant, and $r_{\rm vir}$ is the 
corresponding virial radius. The corresponding virial velocity is
\begin{equation}
  V^2_{\rm vir}  =  \frac{GM_{\rm vir}}{r_{\rm vir}} = (GH)^{2/3} \left(\frac{\Delta_{\rm vir}}{2}\right)^{1/3} M_{\rm vir}^{2/3},
\end{equation}
which can be written more compactly as
\begin{equation}
  V_{\rm vir}  \simeq 35 \left(\frac{\Delta_{\rm vir}}{200}\right)^{1/6} \left(\frac{M_{\rm vir}}{10^{10}h^{-1}\rm M_{\odot}}\right)^{1/3}\,f(z)^{1/2}\mathrm{km/s},
\end{equation}
where 
\begin{equation}
  f(z)=\Omega_M(1+z)^3+\Omega_{\Lambda}
\end{equation}
takes into account evolution where redshift $z$, and $\Omega_M$ and $\Omega_{\Lambda}$ are
the matter and dark energy density parameters respectively.
We assume that the halo contains a reservoir of hot gas whose mass is 
approximately $M_{\rm gas}\sim f_b M_{\rm vir}$, where $f_{\rm b}=\Omega_{\rm b}/\Omega_{\rm M}$ is 
the cosmic mean baryon fraction and $\Omega_{\rm b}$ is the 
baryon density parameter. This is an upper limit -- \citet{crain.etal.2007} 
found that the baryon fraction within $r_{\rm vir}$ is approximately 90 per cent of the cosmic 
mean, independent of $M_{\rm vir}$ and $z$, in cosmological non-radiative hydrodynamical 
simulations, while galaxy formation simulations that account explicitly for the presence of a 
background radiation at high redshifts suggest baryon fractions closer to 10 per cent 
\citep[cf. Figure 3][and references therein]{2014MNRAS.440...50M}. We use the virial 
temperature $T_{\rm vir}$ of the 
halo, 
\begin{equation}
  T_{\rm vir} = \frac{\mu\,m_p}{2k} V_{\rm vir}^2,
\end{equation}
which we approximate as
\begin{equation}
  T_{\rm vir} \simeq 4.4 \times 10^4 \left(\frac{\Delta_{\rm vir}}{200}\right)^{1/3} \left(\frac{M_{\rm vir}}{10^{10}h^{-1}\rm M_{\odot}}\right)^{2/3}\,f(z)\,\mathrm{K}.
\end{equation}
From this we compute the threshold halo mass equivalent to the $T_{\rm vir}$ required for 
efficient cooling, the prerequisite for star formation to occur. We consider two values : 
$T_{\rm vir} \simeq 10^3$ K and $\simeq 10^4$ K. For haloes with $T_{\rm vir} \sim 10^4$ K and
below, primordial gas can cool efficiently through collisional excitation of atomic hydrogen
\citep[see, for example, ][]{moore.etal.2006}. However, molecular hydrogen can also act as an 
efficient coolant in haloes with $T_{\rm vir} \gtrsim 10^3$ K, and it's likely that this
was the dominant cooling mechanism in the earliest generations of star forming haloes  
\citep[e.g.][]{glover.2005}. 

As gas cools into the centre of the halo, we might expect it to form a disc-like structure at 
the centre of the potential -- assuming the system to be dynamically relaxed -- whose radial
extent is governed by the initial angular momentum of the gas 
\citep[cf.][]{1980MNRAS.193..189F,1998MNRAS.295..319M} 
and vertical extent is set by its temperature. If the gas within the disc is in hydrostatic 
equilibrium, the condition for a thin disc to form is that the circular velocity $V_{\rm circ}$ is 
much greater than the sound speed of the gas, 
\begin{equation}
  c_{\rm snd} \simeq \sqrt{\frac{k\,T}{m_{\rm p}}},
\end{equation}
where $T$ is the temperature of the gas, $k$ is the Boltzmann constant and $m_{\rm p}$ 
is the proton mass. Gas temperatures of $10^4/10^3/10^2$K correspond to 
sound speeds of $\sim 10/3/1$ km/s. In these cases, $c_{\rm snd}/V_{\rm circ} \sim 1$ 
implies halo masses ranging from $\sim 3.5 \times 10^8 {\rm M}_{\odot}$ for 10 km/s 
to $10^7 {\rm M}_{\odot}$ for 3 km/s to $\sim 3.5 \times 10^5 {\rm M}_{\odot}$ for 1 km/s. 
In this limit 
discs are likely to be diffuse, vertically 
extended structures 
with low surface densities, and so we reason that star formation is likely to be 
inefficient. We note that a similar conclusion has been reached by 
\citet{2014MNRAS.440...50M}, who investigated the structure and stability of gas discs
in dwarf spheroidal progenitors (see their \S~4.2).

The maximum rate at which gas accumulates in the disc to form stars will be dynamical,
\begin{equation}
  \dot{M} = f_g \frac{M_{\rm vir}}{t_{\rm dyn}} \simeq 0.1 \sqrt{\frac{2}{\Delta_{\rm vir}}} \left(\frac{f_g}{0.1}\right) \left(\frac{M_{\rm vir}}{10^{10}{\rm M}_{\odot}}\right) {\rm M}_{\odot}/{\rm yr},
\end{equation}
which implies that at most $1.5 \times 10^6 {\rm M}_{\odot}$ of gas could settle in the inner 
parts of a $1.5 \times 10^8 {\rm M}_{\odot}$ halo over 100 Myrs. If the star formation 
efficiency was 100 percent, then we would expect 0.2 per cent ($\sim\!2000$, assuming 
a \citet{salpeter.1955} IMF) of the stars that form to have masses in excess of 8 ${\rm M}_{\odot}$,
the threshold mass for the star to go supernova. However, we expect star formation efficiencies much lower than this.

Interestingly, the timescale for star formation can be an appreciable fraction of the
age of the Universe at high redshifts. Stars form from cold dense clouds of gas, which
must collapse sufficiently before a star can reach the zero age Main Sequence (ZAMS), 
and the timescale for this to occur is much longer for low-mass stars than their high 
mass counterparts. 
The pre-main sequence lifetime (i.e. the timescale to reach ZAMS) of a star of mass $m_{\ast}$
can be approximated as
\begin{equation}
t_{\rm PMS} \simeq 10^7\, \left(\frac{m_{\ast}}{\rm M_{\odot}}\right)^{-2.5} {\rm yr}
\label{eq:tpms}
\end{equation}
\noindent and so massive stars reach ZAMS much earlier than lower mass stars 
\citep[cf.][]{2013sse..book.....K} . A 
1 $\rm M_{\odot}$ star reaches ZAMS after $10^7$yr, whereas an 8 $\rm M_{\odot}$ 
reaches ZAMS after $\sim 5.5 \times 10^4$ yr, which implies that high mass stars can
reach ZAMS and evolve off the MS as supernovae, each depositing $10^{51}$ erg of 
energy into their surroundings, before lower mass stars have had time to form. For 
example, a 20 $\rm M_{\odot}$ star will have had sufficient time to reach ZAMS and evolve
off the Main Sequence ($t_{\rm MS} \sim 10^7$ yr) before a 1 $\rm M_{\odot}$ star has reached
ZAMS.

Finally, we compare the amount of energy liberated by supernovae and deposited into the 
surrounding gas to the binding energy of gas in a dark matter halo. For a 
NFW profile, the binding energy is

\begin{equation}
  W = W_0 \times \left[\frac{c}{1+c}-\frac{\log(1+c)}{(1+c)}\right]
\end{equation}

\noindent where $c=r_{\rm vir}/r_s$ is the halo's concentration with $r_s$ the 
scale radius, and

\begin{equation}
  W_0 = -\frac{16\pi^2}{2}f_b G\rho_{\rm crit}^2\left(\frac{\Delta_{\rm vir}}{3}\frac{c^3}{\log(1+c)-c/(1+c)}\right)^2\left(\frac{r_{\rm vir}}{c}\right)^5.
\end{equation}

\noindent We can simplify this to get

\begin{equation}
  W \simeq 10^{56} \left(\frac{M_{\rm vir}}{10^{10} {\rm M}_{\odot}/h}\right)^{5/3} \left(\frac{f_b}{0.16}\right) \mathcal{F}(c) {\rm erg}
\end{equation}

\noindent where

\begin{equation}
  \mathcal{F}(c) = c \times \frac{\left[c/(1+c)-\log(1+c)/(1+c)\right]}{\left[\log(1+c)-c/(1+c)\right]^2};
\end{equation}

\noindent for reasonable values of $c$, $\mathcal{F}(c) \sim 3-4$. For a 
$10^7 {\rm M}_{\odot}$ halo, the binding energy of gas is of order the 
mechanical energy of a supernova. We have argued that the gaseous discs in these mass
systems are likely to be diffuse and extended vertically, and so it is likely to be relatively
easily disrupted by the impact of supernovae (and preceeding photo-ionization and stellar winds
by their Main Sequence progenitors).

\section{Star Formation in Low-Mass Haloes: Monte Carlo Modelling}
\label{sec:model}

\subsection{Assembly Histories}
\label{subsec:assembly_histories}

We have used the algorithm of \citet{parkinson.etal.2008} to generate Monte 
Carlo merger histories -- so called trees -- for dark matter haloes with masses 
between $M_{\rm vir}$=$10^7 {\rm M}_{\odot}$ to $10^{10} {\rm M}_{\odot}$ at $z$=0, in 
the range we would expect to bracket the virial masses of the hosts of present day 
satellite galaxies. These merger trees track the evolution of the haloes identified with
mass $M$ at $z$=0 to an initial redshift of $z$=30, which we capture at uniform intervals 
in the logarithm of the expansion factor, $a=1/(1+z)$. Because this particular 
implementation has been calibrated against $N$-body trees drawn the Millennium simulation, 
we can rapidly explore a range of realistic merger histories for a statistical sample of 
haloes. For this reason, we adopt the cosmological parameters used for the Millennium 
Simulation ($\Omega_{\rm 0}$=0.25, $\Omega_{\rm 0}$=0.75, $h$=0.73 and $\sigma_8$=0.9) when 
constructing these trees, rather than the more recent Planck parameters 
\citep[cf.][]{2013arXiv1303.5076P}; the differences will be quantitative rather than
qualitative, driven by the change in $\sigma_8$, and will not affect the logic of our
argument.

\begin{figure}
  \centering
  \includegraphics[width=\columnwidth]{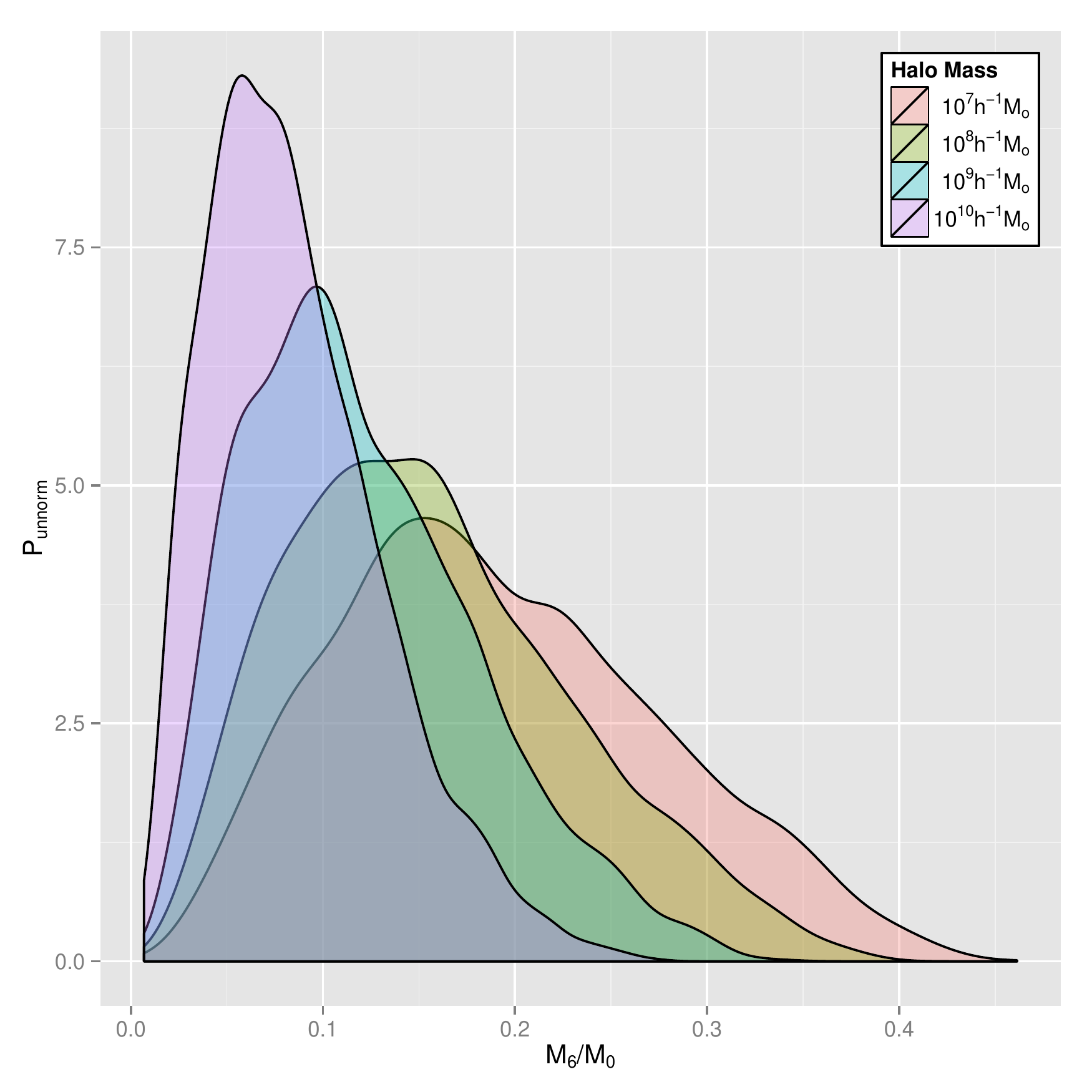}
  \caption{{\bf Satellite Halo Growth since $z$=6.} We plot the unnormalised probability
    distributions of $\mathcal{M}_{z=6}=M(z=6)/M(z=0)$ for haloes with $z$=0 masses of
    $10^7 {\rm M}_{\odot}$, $10^8 {\rm M}_{\odot}$, $10^9 {\rm M}_{\odot}$ and $10^{10} {\rm M}_{\odot}$ 
    (red, green, blue and purple filled regions respectively).
  }
  \label{fig:mgrowth_z6}
\end{figure}

We begin by using the trees to explore the factors by which haloes of present day
virial mass $M$ have grown since a fixed redshift in the early Universe.
For each halo mass we sample 5000 trees and for each halo tree we determine
progenitor masses at $z$=6 and $z$=10. Median and upper lower quartiles of the 
ratio $M(z)/M(z=0)$ are summarised in Table~\ref{tab:assembly_histories}, while the distributions
of these values for halo growth since $z$=6 are shown in Figures~\ref{fig:mgrowth_z6} for present-day
halo masses of $5 \times 10^7$, $10^8$, $10^9$ and $10^{10} {\rm M}_{\odot}$. As a 
general rule, the more massive the halo, the more rapidly it has grown since high redshift 
-- e.g. a $5 \times 10^7 {\rm M}_{\odot}$ ($10^{10} {\rm M}_{\odot}$) halo at $z$=0 has grown 
by a median factor of $\sim 5$ ($\sim 10$) since $z$=6 and $\sim 17$ ($\sim 62$) since $z$=10 --
but also the narrow the distribution of $M(z)/M(z=0)$, the factor by which it has grown in mass between 
$z$ and $z$=0.

\subsection{Epoch of Initial Star Formation}

\begin{figure}
  \centering
  \includegraphics[width=\columnwidth]{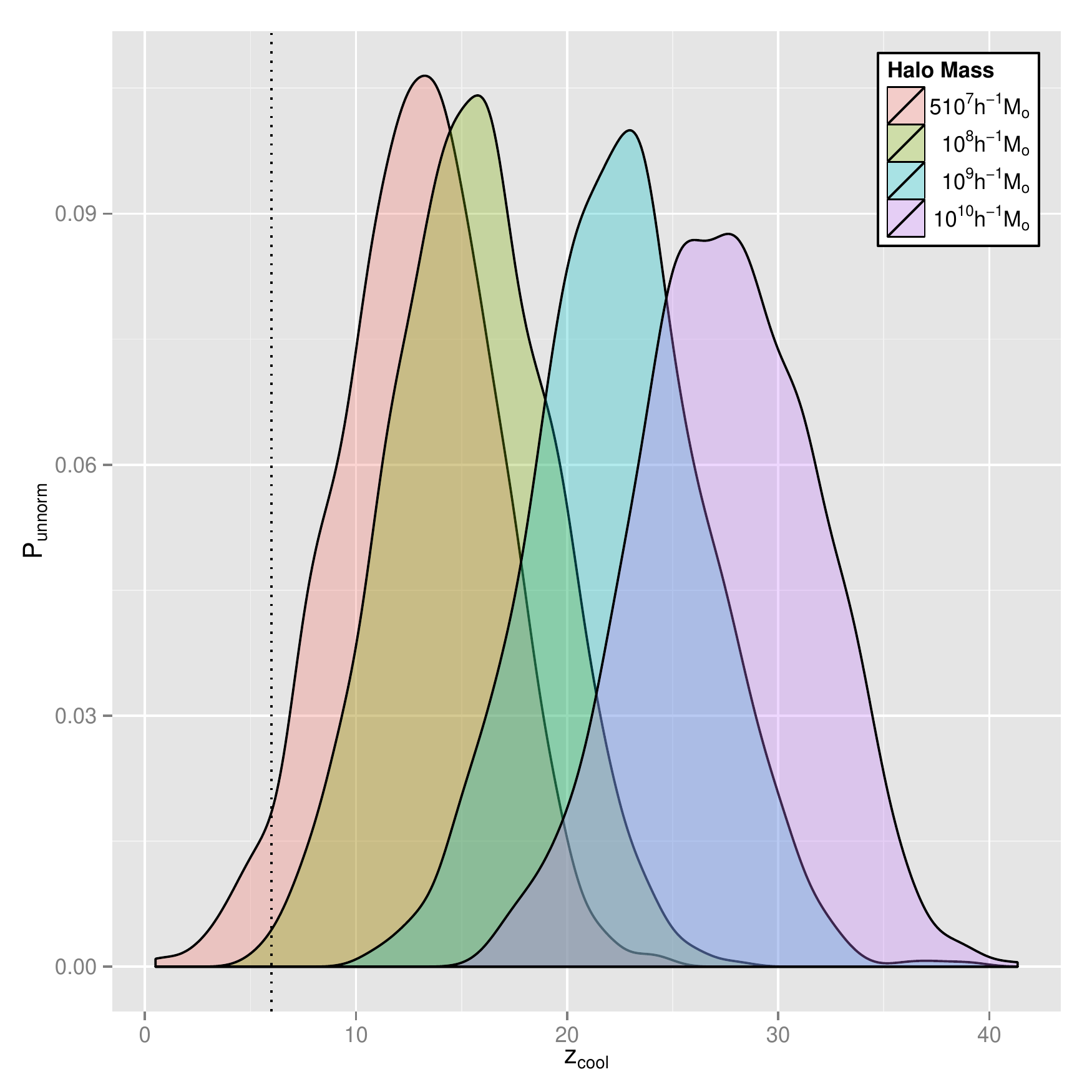}
  \caption{{\bf Redshift at which Cooling is Supported.}  Here 
    we plot the unnormalised probability distributions of $z_{\rm cool}$ for 
    haloes with $z$=0 masses of
    $10^9 {\rm M}_{\odot}$ and $10^{10} {\rm M}_{\odot}$ 
    (blue and red filled regions respectively).
  }
  \label{fig:zcool}
\end{figure}

For each halo we determine $z_{\rm cool}$, the redshift at which a progenitor halo becomes 
sufficiently massive to support cooling and consequently star formation. In 
Figure~\ref{fig:zcool} we show the distribution of $z_{\rm cool}$ for haloes with present
day masses between $10^7 {\rm M}_{\odot}$ and $10^{10} {\rm M}_{\odot}$ first exceed a virial 
temperature of $10^3$K, the threshold for cooling via molecular hydrogen; haloes less massive 
than $\sim 10^9$K never become sufficiently massive to support cooling via atomic hydrogen. 
The key point to note in this Figure is that more massive haloes can support cooling earlier 
in their assembly histories. However, this assumes that haloes form in regions of mean cosmic 
density. We can mimic the effect of haloes forming in overdense environments -- which is likely 
for lower mass haloes -- by increasing the value of $\sigma_8$ \citep[e.g.][]{boley.etal.2009},
which shifts $z_{\rm cool}$ to high redshifts. For example, increasing $\sigma_8$ from 0.8 to 1.0 
increases $z_{\rm cool}$ from $\sim 15.5$ to $\sim 21.3$, a difference of $\approx 10^8$ years
in our adopted cosmology.

For each of these haloes, we mimic the effect of inefficient star formation by assuming that 
only a fraction $f_{\ast}$ of the gas content of the halo at $z_{\rm cool}$ can cool and form 
stars. We consider cases in which (i) $f_{\ast}$ is the same for all haloes and (ii) $f_{\ast}$ is
drawn from a Gaussian distribution, with a mean and standard deviation that we are free to choose;
$f_{\ast}$ encapsulates our ignornance about the physics that drives star formation and 
so acts as a formation efficiency. From this star-forming gas mass, 
\begin{equation}
  M_{\ast}=f_{\ast}f_{\rm b}M_{\rm vir}(z_{\rm cool}),
\end{equation}
we sample an initial mass function $\xi(m)$ to obtain the distribution of stellar masses and 
count the number of stars with masses in excess of 8 M$_{\odot}$. Here we use the \citet{salpeter.1955} 
IMF (i.e. $N(m) \propto m^{-1.35}$), but we have verified that the choice
of IMF does not influence our results by using also the \citet{kroupa.2001} and \citet{chabrier.2001} 
IMFs. Note that the larger the value we choose for $f_{\ast}$, the greater the number of massive stars that 
form and the larger the energy they can deposit into the interstellar and circumgalactic medium (ISM 
and CGM respectively). We consider values of $f_{\ast}$ that vary between $10^{-6}$ and $10^{-3}$; the 
effect is linear in $f_{\ast}$, as we would expect, and so we are more interested in relative rather than
absolute values versus $M$ and $z_{\rm cool}$.

\begin{figure}
  \centering
  \includegraphics[width=\columnwidth]{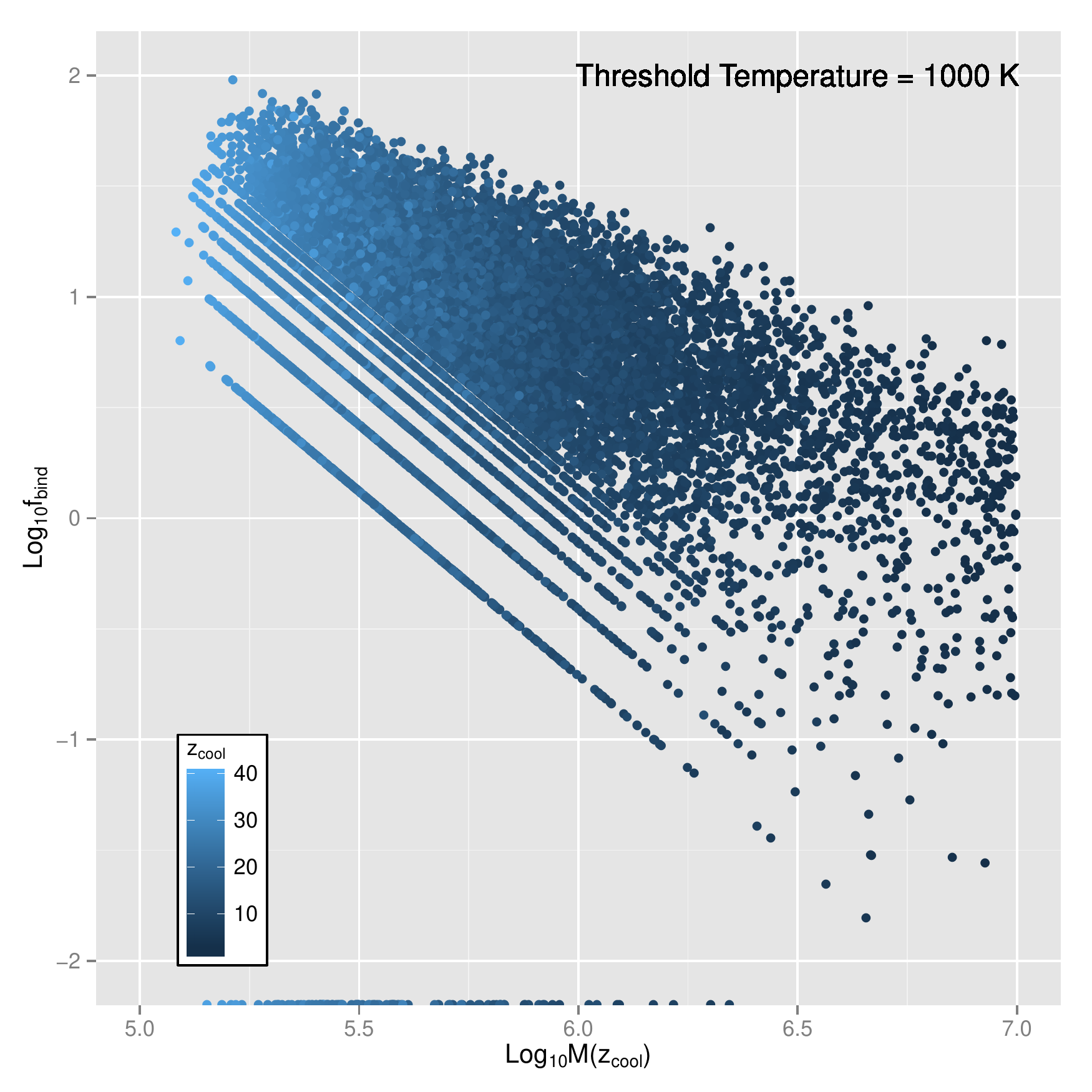}
  \caption{{\bf Fraction of Binding Energy Deposited by Supernovae as a
      function of halo mass at $z_{\rm cool}$}. Points are coloured by
      $z_{\rm cool}$, as indiciated by the colour bar.}
  \label{fig:funbind}
\end{figure}

The fate of gas in these nascent galaxies is governed by the high mass stars that form, each 
depositing $\sim\!10^{51}$ ergs into the ISM and CGM, which can unbind gas in the halo and 
possibly prevent further stars from forming. We compute the ratio of supernovae-deposited 
energy to the binding energy of the proto-galaxy as soon as the first massive stars have had 
sufficient time to form and evolve off the main sequence 
($\Delta t \sim t_{\rm PMS} + t_{\rm MS}$; cf. Eq~\ref{eq:tpms}) subsequent to 
the halo becoming sufficiently massive to support cooling. In Figure~\ref{fig:funbind} we plot 
the ratio $f_{\rm bind}=E_{\rm SNe}/E_{\rm bind}$ against halo mass at $z_{\rm cool}$, where we 
compute $f_{\rm bind}$ from energy deposited by supernovae whose progenitor initial masses were in excess of 8 M$_{\odot}$ and, recall, we identify $z_{\rm cool}$ as the redshift at 
which the virial temperature of the halo first exceeds $10^3$K. 

\begin{figure}
  \centering
  \includegraphics[width=\columnwidth]{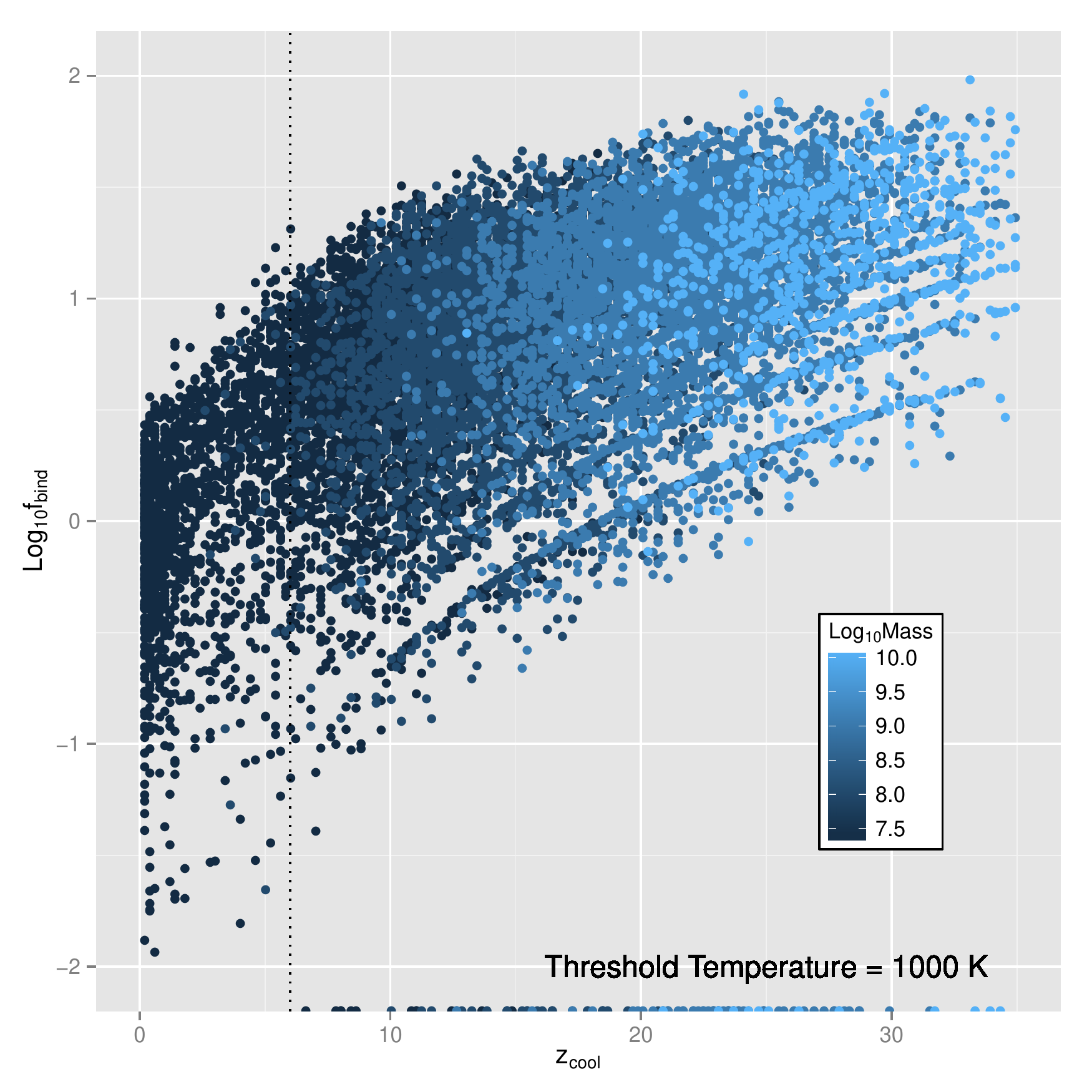}
  \caption{{\bf Fraction of Binding Energy Deposited by Supernovae as a
      function of $z_{\rm cool}$}, for haloes with present-day masses between
      $5 \times 10^7$ and $10^{10}$ $\rm M_{\odot}$. Points are coloured by
      the present-day halo mass, as indiciated by the colour bar. Vertical
      dotted line indicates the redshift at which we expect reionization to 
      be complete, which we use as a natural boundary for this process to be
      viable.}
  \label{fig:funbind_zcool}
\end{figure}

As we would expect, Figure~\ref{fig:funbind} shows that supernovae are most destructive in the 
haloes that have the lowest masses -- and therefore the shallowest potentials -- at 
$z_{\rm cool}$, depositing sufficient energy to unbind the gas in the halo. However, inspecting
the key, which shows how the points are colour-coded in accord with their $z_{\rm cool}$, is 
instructive; these low-mass systems are more likely to be the high redshift progenitors of 
present-day higher mass haloes ($M \gtrsim 10^9 \rm M_{\odot}$). These haloes will continue to 
grow and they can potentially reaccrete gas and support atomic line cooling once their virial 
temperatures exceed $10^4$K; depending on their final mass, this bursty mode of star formation 
may be traceable in the chemistry of their stars and accessible to observations
\citep[e.g.][]{2013ApJ...779..102K} . In contrast, the progenitors of present-day
low-mass haloes are more likely to reach the threshold for forming their stars later 
(i.e. lower $z_{\rm cool}$) when 
they are more massive and so supernovae will less effective, a point that is made clear in
Figure~\ref{fig:funbind_zcool}. However, the late onset of cooling in these systems means that
it is more likely that the ambient radiation field in which these systems reside will reionize
the gas that would otherwise collapse and cool in these haloes. For this reason, the effects
of stochastic star formation and feedback are likely to be most pronounced in haloes with
present-day masses of $\sim 10^9-10^{10} \rm M_{\odot}$.

\begin{table}
\begin{center}
  \caption{\textbf{Satellite Halo Assembly Histories.} $M$($z$=0) corresponds to the satellite
    halo mass at $z$=0; $\mathcal{M}_{z=6}$ and $\mathcal{M}_{z=10}$ correspond to the median 
    value of the ratio of the $z$=6 and 10 progenitor masses to the $z$=0 mass, with associated 
    upper and lower quartiles; and $z_{\rm cool}$ represents the median redshift, with upper and 
    lower quartiles, at which the $z$=0 halo's progenitors were sufficiently massive to support 
    cooling by molecular ($T_{\rm vir}=10^3 \rm K$, first value) and atomic 
    ($T_{\rm vir}=10^4 \rm K$, second value) hydrogen respectively.}
\vspace*{0.3 cm}

\begin{tabular}{lccc}\hline
$M$ [$h^{-1} {\rm M}_{\odot}$]  & $\mathcal{M}_{z=6} $& $\mathcal{M}_{z=10}$ & $z_{\rm cool}$ \\
\hline
$5 \times 10^7$    &  $0.18^{+0.07}_{-0.05}$  & $0.058^{+0.028}_{-0.021}$ & $12.8^{+2.8}_{-2.4}$ / ~~~- \\
$10^8$    &  $0.15^{+0.05}_{-0.05}$  & $0.041^{+0.021}_{-0.015}$ & $15.5^{+2.4}_{-2.6}$ / ~~~- \\
$10^9$    &  $0.11^{+0.05}_{-0.03}$  & $0.027^{+0.015}_{-0.011}$ & $22.3^{+2.8}_{-2.8}$ / ~$8.2^{+2.0}_{-2.4}$ \\
$10^{10}$ & $0.08^{+0.03}_{-0.03}$ & $0.016^{+0.009}_{-0.006}$ & $27.7^{+2.8}_{-3.0}$ / $14.6^{+2.4}_{-2.0}$ \\

\hline
\end{tabular}
\label{tab:assembly_histories}
\end{center}
\end{table}

\section{Summary}
\label{sec:summary}

Using simple physical arguments, we explored star formation and associated 
feedback in low-mass dark matter haloes. We reasoned that star formation 
should proceed in an inefficient fashion in these systems and consequently that the 
formation rate of high mass stars ($m_{\ast} \geq 8 \rm M_{\odot}$) and resulting 
supernovae\footnote{Although we focus on supernovae, this applies equally to other forms
of feedback by Main Sequence progenitors, namely photo-ionization and stellar winds.} is 
likely to be stochastic. Because the star formation timescale decreases sharply with 
increasing stellar mass, as $t_{\rm PMS} \propto m_{\ast}^{-2.5}$, we argued that high 
mass stars that form early can both disrupt subsequent lower mass star formation and 
drive gas, which would otherwise be available for star formation, out of low-mass haloes.

Using Monte Carlo assembly histories of low mass haloes with present-day masses 
$10^7 {\rm M}_{\odot} \lesssim M \leq 10^{10} {\rm M}_{\odot}$, we estimated when a
given halo mass was likely to have first supported cooling and star formation by
determining the redshift at which the progenitor virial temperature first exceeded
thresholds of $10^3$K (for molecular hydrogen cooling) and $10^4$K (for atomic 
hydrogen cooling). Haloes with present-day masses $\lesssim 10^9 \rm M_{\odot}$ 
tend to reach the threshold for cooling via molecular hydrogen later. If they reach
the threshold prior to the onset of cosmological reionization and form high mass stars, 
their potential wells are sufficiently shallow that gas can be expelled by supernovae, 
they never become massive enough to re-accrete gas that cools, and the expelled gas is 
likely to be further heated by cosmological reionization; otherwise, if they don't reach the
threshold prior to reionization, gas can never cool and star formation cannot proceed.
The epoch of star formation in these systems is likely to be concentrated in a single burst,
and may explain observations such as those of \citet{2012ApJ...753L..21B}, who examined the 
star formation histories of ultra-faint dwarf galaxies, the least luminous and most 
dark matter dominated systems known, and found no evidence for extended star formation 
histories. In contrast, haloes with present-day masses $\sim\!10^9$ to $10^{10}\rm M_{\odot}$ 
reach the 
threshold for cooling via molecular hydrogen cooling early ($z \gtrsim 20$) and, while 
supernovae may disrupt star formation and expel gas, the halo grows sufficiently to 
reach the threshold for atomic hydrogen cooling and there is time for gas to be re-accreted.
These systems are likely to experience episodic star formation, which may be inferred
from the chemistry of their stellar populations. 

In this way, inefficient star formation in low-mass haloes and the disparity in timescales 
for the formation and of low- and high-mass stars, with their associated feedback, conspires
to produce an internal mechanism for quenching galaxy formation in low-mass haloes in a 
stochastic fashion. This will act in conjunction with external mechanisms, such as the ionizing 
radiation background that was present during cosmological reionization 
\citep[e.g.][]{weinmann.etal.2007,lunnan.etal.2012} . This is in accord
with the adaptive mesh refinement cosmological zoom simulation of \citet{2013MNRAS.432.1989S},
who studied the formation and evolution of a galaxy in a $10^9 \rm M_{\odot}$ dark matter halo 
at high redshift. They found that cosmological reionization is responsible for expelling
most of the low density gas from the halo, while supernovae are required to disperse
the cool, high density gas from the disc.

We have presented a general argument, but we note that it provides 
a plausible explanation for the implied low and apparently stochastic rate of occupancy of 
satellite galaxies in dark matter haloes with masses 
$M \lesssim 10^{10} {\rm M}_{\odot}$, which appears to be required to reconcile observations
of satellite galaxies around the Milky Way with dark matter subhaloes in MW-type systems
in cosmological $N$-body simulations \citep[cf.][]{boylan.etal.2011}. 

We have focussed on the effects of high mass star formation and supernovae, 
but we note also that stars more massive than 2.5 ${\rm M}_{\odot}$ form in 
multiplets, and can evolve to form X-ray binaries. These can be extremely 
luminous, both radiatively and mechanically, and unbind gas in the galaxy 
\citep[e.g.][]{2012MNRAS.423.1641J,power.etal.2013}.
At the same time, their formation depends on details of the formation of stellar 
multiplets and binary evolution, factors that are for all intents and purposes 
stochastic.

Finally we note that this process should not depend on the underlying dark 
matter model -- the nature of the physical processes (e.g. cooling, star 
formation, feedback) will not and so galaxy formation in low-mass haloes will 
be subject to the same physically-imposed limitations across dark matter models 
\citep[see discussion in, for example,][]{power.2013}. 

\section*{Acknowledgments}
CP thanks the GALFORM Team for making the merger tree algorithm of \citet{parkinson.etal.2008} 
publically available. CP and GFL acknowledge support of Australian Research Council (ARC) 
DP130100117; CP, ASGR, and GFL acknowledge support of ARC DP140100198. CP acknowledges 
support of ARC FT130100041. The research presented in this paper is undertaken as part of 
the Survey Simulation Pipeline (SSimPL; {\texttt{http://ssimpl.org/}).

\bibliographystyle{mn2e}

\label{lastpage}

\end{document}